\documentclass[twoside,twocolumn,9pt]{article}
\usepackage{extsizes}
\usepackage[super,sort&compress,comma]{natbib} 
\usepackage[version=3]{mhchem}
\usepackage[left=1.5cm, right=1.5cm, top=1.785cm, bottom=2.0cm]{geometry}
\usepackage{balance}
\usepackage{mathptmx}
\usepackage{sectsty}
\usepackage{graphicx} 
\usepackage{lastpage}
\usepackage[format=plain,justification=justified,singlelinecheck=false,font={stretch=1.125,small,sf},labelfont=bf,labelsep=space]{caption}
\usepackage{float}
\usepackage{fancyhdr}
\usepackage{fnpos}
\usepackage[english]{babel}
\addto{\captionsenglish}{%
  
}
\usepackage{array}
\usepackage{droidsans}
\usepackage{charter}
\usepackage[T1]{fontenc}
\usepackage[usenames,dvipsnames]{xcolor}
\usepackage{setspace}
\usepackage[compact]{titlesec}
\usepackage{hyperref}
\usepackage{gensymb}
\usepackage{makecell}

\usepackage{placeins}

\usepackage{epstopdf}

\definecolor{cream}{RGB}{222,217,201}

\begin{document}

\pagestyle{fancy}
\thispagestyle{plain}
\fancypagestyle{plain}{
\renewcommand{\headrulewidth}{0pt}
}

\makeFNbottom
\makeatletter
\renewcommand\LARGE{\@setfontsize\LARGE{15pt}{17}}
\renewcommand\Large{\@setfontsize\Large{12pt}{14}}
\renewcommand\large{\@setfontsize\large{10pt}{12}}
\renewcommand\footnotesize{\@setfontsize\footnotesize{7pt}{10}}
\makeatother

\renewcommand{\thefootnote}{\fnsymbol{footnote}}
\renewcommand\footnoterule{\vspace*{1pt}%
\color{cream}\hrule width 3.5in height 0.4pt \color{black}\vspace*{5pt}} 
\setcounter{secnumdepth}{5}

\makeatletter 
\renewcommand\@biblabel[1]{#1}            
\renewcommand\@makefntext[1]%
{\noindent\makebox[0pt][r]{\@thefnmark\,}#1}
\makeatother 
\renewcommand{\figurename}{\small{Fig.}~}
\sectionfont{\sffamily\Large}
\subsectionfont{\normalsize}
\subsubsectionfont{\bf}
\setstretch{1.125} 
\setlength{\skip\footins}{0.8cm}
\setlength{\footnotesep}{0.25cm}
\setlength{\jot}{10pt}
\titlespacing*{\section}{0pt}{4pt}{4pt}
\titlespacing*{\subsection}{0pt}{15pt}{1pt}

\fancyfoot{}
\fancyfoot[LO,RE]{\vspace{-7.1pt}\includegraphics[height=9pt]{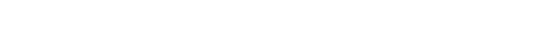}}
\fancyfoot[CO]{\vspace{-7.1pt}\hspace{11.9cm}\includegraphics{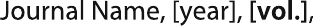}}
\fancyfoot[CE]{\vspace{-7.2pt}\hspace{-13.2cm}\includegraphics{RF}}
\fancyfoot[RO]{\footnotesize{\sffamily{1--\pageref{LastPage} ~\textbar  \hspace{2pt}\thepage}}}
\fancyfoot[LE]{\footnotesize{\sffamily{\thepage~\textbar\hspace{4.65cm} 1--\pageref{LastPage}}}}
\fancyhead{}
\renewcommand{\headrulewidth}{0pt} 
\renewcommand{\footrulewidth}{0pt}
\setlength{\arrayrulewidth}{1pt}
\setlength{\columnsep}{6.5mm}
\setlength\bibsep{1pt}

\makeatletter 
\newlength{\figrulesep} 
\setlength{\figrulesep}{0.5\textfloatsep} 

\newcommand{\topfigrule}{\vspace*{-1pt}%
\noindent{\color{cream}\rule[-\figrulesep]{\columnwidth}{1.5pt}} }

\newcommand{\botfigrule}{\vspace*{-2pt}%
\noindent{\color{cream}\rule[\figrulesep]{\columnwidth}{1.5pt}} }

\newcommand{\dblfigrule}{\vspace*{-1pt}%
\noindent{\color{cream}\rule[-\figrulesep]{\textwidth}{1.5pt}} }

\makeatother

\twocolumn[
  \begin{@twocolumnfalse}
{\includegraphics[height=30pt]{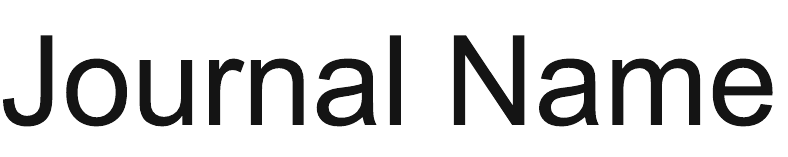}\hfill\raisebox{0pt}[0pt][0pt]{\includegraphics[height=55pt]{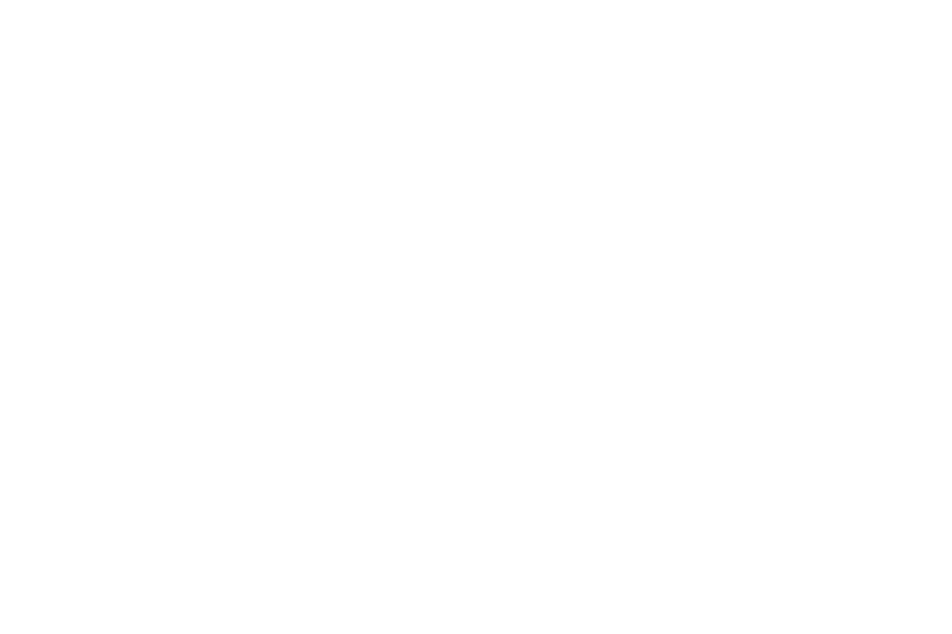}}\\[1ex]
\includegraphics[width=18.5cm]{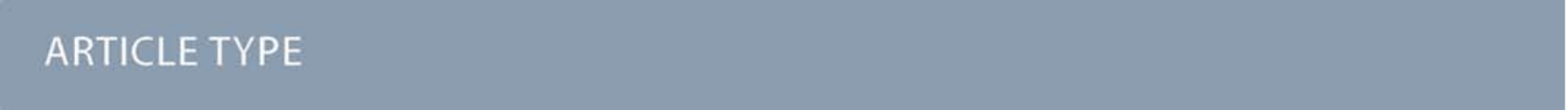}}\par
\vspace{1em}
\sffamily
\begin{tabular}{m{4.5cm} p{13.5cm} }

\includegraphics{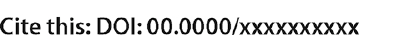} & \noindent\LARGE{\textbf{Discovery of Superconductivity in \ce{Nb4SiSb2} with a \ce{V4SiSb2}-Type Structure and Implications of Interstitial Doping on its Physical Properties}} \\%
\vspace{0.3cm} & \vspace{0.3cm} \\

 & \noindent\large{Manuele D. Balestra,\textit{$^{ab}$} Omargeldi Atanov \textit{$^{c}$}, Robin Lefèvre,\textit{$^{b}$} Olivier Blacque,\textit{$^{b}$} Yat Hei Ng,\textit{$^{c}$} Rolf Lortz,\textit{$^{c}$} Fabian O. von Rohr\textit{$^{a\ast}$}} \\


\includegraphics{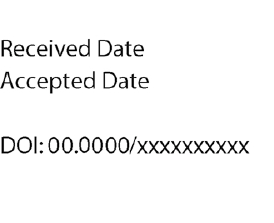} & \noindent\normalsize{

We report on the discovery, structural analysis, and the physical properties of \ce{Nb4SiSb2} -- a \textit{hitherto} unknown compound crystallizing in the \ce{V4SiSb2}-type structure with the tetragonal space group \textit{I}4\textit{/mcm} and unit cell parameters \textit{a}\, = 10.3638(2)\, \AA~ and \textit{c}\, = 4.9151(2)\, \AA. We find \ce{Nb4SiSb2} to be a metal undergoing a transition to a superconducting state at a critical temperature of $T_{\rm c} \approx$ 1.6 K. The bulk nature of the superconductivity in this material is confirmed by the observation of a well defined discontinuity in specific heat with a normalized specific heat jump of $\Delta C(T_{\rm c})/\gamma T_{\rm c} = 1.33\, {\rm mJ}\, {\rm mol}^{-1}\, {\rm K}^{-2}$. We find that for \ce{Nb4SiSb2}, the unoccupied sites on the 4\textit{b} \textit{Wyckoff} position can be partially occupied with Cu, Pd, or Pt. Low-temperature resistivity measurements show transitions to superconductivity for all three compounds at $T_{\rm c} \approx\, 1.2\, {\rm K}$ for \ce{Nb4Cu_{0.2}SiSb2}, and $T_{\rm c} \approx\, 0.8\, {\rm K}$ for \ce{Nb4Pd_{0.2}SiSb2} as well as for \ce{Nb4Pt_{0.14}SiSb2}. The addition of electron-donor atoms into these void positions, henceforth, lowers the superconducting transition temperature in comparison to the parent compound. 

} \\

\end{tabular}

 \end{@twocolumnfalse} \vspace{0.6cm}

  ]

\renewcommand*\rmdefault{bch}\normalfont\upshape
\rmfamily
\section*{}
\vspace{-1cm}


\footnotetext{\textit{$^{a}$~Department of Quantum Matter Physics, University of Geneva, CH-1211 Geneva, Switzerland.}}
\footnotetext{\textit{$^{b}$~Department of Chemistry, University of Zurich, CH-8057 Zurich, Switzerland.}}
\footnotetext{\textit{$^{c}$~Department of Physics, The Hong Kong University of Science and Technology, Clear Water Bay Kowloon, Hong Kong.}}
\footnotetext{\textit{$^{\ast}$~ E-mail to the author}}





\section{Introduction}
\label{Introduction}

\begin{figure*}[h]
 \centering
 \includegraphics[height=12cm]{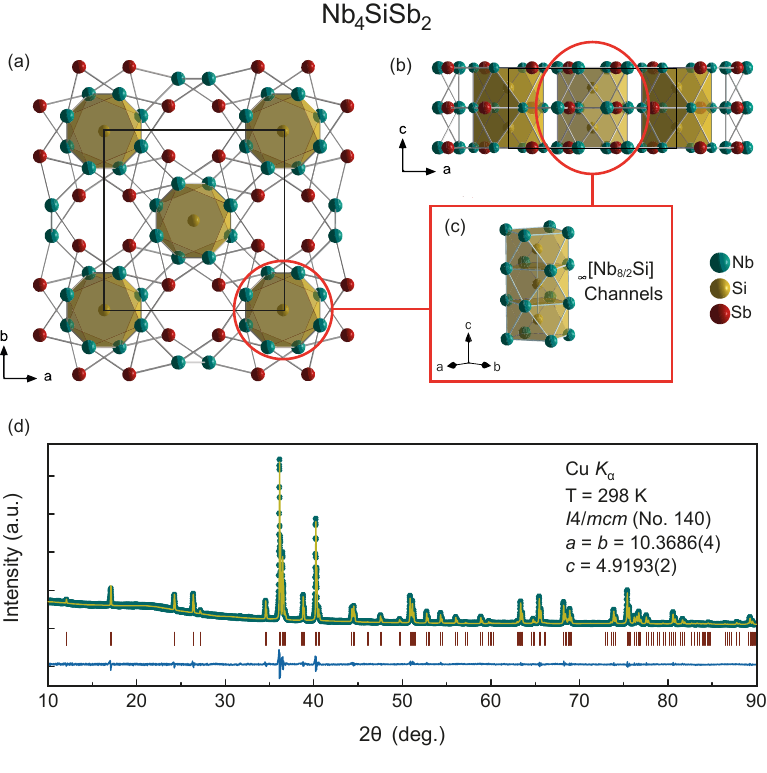}
 \caption{Schematic representation of crystal structure obtained from SXRD refinement of \ce{Nb4SiSb2} along (a) the \textit{c}-direction and (b) the \textit{b}-direction. (c) Si\, --\, Si chains along the \textit{c}-direction. (d) PXRD pattern of the polycrystalline sample with the respective \textit{Rietveld} refinement. Green dots: data points, yellow line: calculated peaks, vertical dark red lines: Bragg peak positions, and the blue pattern on the bottom is the difference plot.}
 \label{fig_nb4sisb2_sc_and_rietveld}
\end{figure*}

A promising approach for the discovery of new superconducting materials is based on the substitution or incorporation of elements into existing structures with crystallographic void positions. Substituting or incorporating atoms into a structure allows for the precise chemical modification of the density of electronic states at the Fermi-level. This may induce superconductivity or tune superconducting properties.\cite{Yazici2013,Zhang2017a,Akimitsu2019,Ma2021_2,Wang2019} Recent examples of this include the increase in the superconducting transition temperature of \ce{Nb5Ge3} -- in the tetragonal \ce{Cr5B3} type-structure -- from $T_{\rm c} \approx 0.7\, {\rm K}$ to 15.3 K by the incorporation of carbon atoms into void positions,\cite{Bortolozo2012} or the stabilization of $\eta$-carbide superconductors in a filled \ce{Ti2Ni}-type structure, with remarkably high upper critical fields.\cite{Ma2021_1, Ma2021_2}

The results presented in this paper refer to structures, crystallizing in a defect variant of the \ce{W5Si3}-type structure, commonly known as the \ce{V4SiSb2} structure. The \ce{W5Si3} structure itself exhibits the tetragonal space group \textit{I}4\textit{/mcm}\cite{Aronsson1955} and is a bulk superconductor with a critical temperature of $T_{\rm c} = 2.7\, K$.\cite{Kawashima2009} Other compounds crystallizing in the same structure-type and exhibiting superconducting properties are \ce{Nb5Si3} with a critical temperature of $T_{\rm c}$ = 0.7\ K\cite{Willis1979}, and the ternary \ce{W5Si3}-type compounds \ce{Nb5Sn2Ga}, \ce{Ta5SnGa2}, and \ce{Zr5Sb_{2.36}Ru_{0.36}} with critical temperatures of $T_{\rm c} \approx$ 1.8 K, 1.8 K, and 5 K, respectively.\cite{Shishido1989,Shishido1989a,Xie2015}.
In the \ce{V4SiSb2} structure, the 4\textit{b} Wyckoff position of the \ce{W5Si3} structure is unoccupied, forming void channels along the \textit{c}-direction. These channels are filled by Sb centred, essentially unhybridized 5p orbitals forming a 2D net stacking along the $c$-direction leading to "electron-filled" voids.\cite{Rytz1999a} The prospect of intercalating these voids with electrophilic species has been theoretically proposed by Rytz \textit{et al.}.\cite{Rytz1999a} 

To date, only six compounds have been reported to crystallize in this structure-type, namely \ce{V4SiSb2} and the compound series of \ce{Ti4\textit{T}Bi2} with \textit{T} = Cr, Mn, Fe, Co, Ni). All of these compounds are known to be non-magnetic metals.\cite{Wollesen1996,Richter1997} Furthermore, 5 pseudo-quaternary antimonides with the general formula \ce{Nb4Pd_{0.5}\textit{Z}Sb2} with Z = Cr, Fe, Co, Ni, Si have been reported.\cite{Wang2001} These compounds contain three transition metals in an ordered arrangement; hence they are isostructural to each other and crystallize in substitutional variants of the \ce{W5Si3}-type structure, or alternatively, they can be interpreted as \ce{V4SiSb2}-type compounds with half occupied channels. 

Here, we report on the discovery of the compound \ce{Nb4SiSb2}, which crystallizes in a \ce{V4SiSb2}-type structure with the tetragonal space group \textit{I}4\textit{/mcm}. We show that this material exhibits bulk superconductivity at a critical temperature of $T_{\rm c} \approx 1.6$ K. Furthermore, we find that the 4\textit{b} Wyckoff void position can be partially occupied by the transition metals Cu, Pd or Pt, leading to the compounds \ce{Nb4Cu_{0.2}SiSb2}, \ce{Nb4Pd_{0.2}SiSb2}, and \ce{Nb4Pt_{0.14}SiSb2}. All three compounds are bulk superconductors with critical temperatures of $T_{\rm c} \approx$ 1.2 K, 0.8 K, and 0.8 K, respectively.

\section{Experimental}
\label{Experimental}

\textit{Synthesis:} Polycrystalline samples of all compounds were obtained by solid state reaction of the pressed elemental powders at high temperatures. These were synthesized using pure elements as received and stored in air of niobium (powder, 99.99\%, \textit{Alfa Aesar}), silicon (pieces, 99.95\%, \textit{Alfa Aesar}), antimony (shots, 99.999\%, \textit{Alfa Aesar}), copper (powder, 99.7\%, \textit{Merck}), palladium (powder, 99.999\%, \textit{Arcos Organics}) and platinum (powder, 99.999\%, \textit{Arcos Organics}). The elements were thoroughly mixed and ground in their stoichiometric ratios, then pressed into pellets, and subsequently sealed in quartz ampoules under 400\, mbar of Ar. The quartz ampoules were heated to $T = 1100\, ^{\circ}$C with a heating rate of 180 $^{\circ}$C/h, and annealed at this temperature for 7 days.\\
\textit{Diffraction:} Single crystal X-ray diffraction (SXRD) data were collected at $T = 160(1)\, K$ on a \textit{Rigaku} XtaLAB Synergy, Dualflex, Pilatus 200K diffractometer using a monochromatic X-ray source (Cu $K_{\alpha_1}$ radiation: $\lambda = 1.54184\,$ \AA) from a micro-focus sealed X-ray tube and cooled using an \textit{Oxford} liquid-nitrogen Cryostream device. The selected suitable single crystals were mounted using polybutene oil. Pre-experiment, data collection, data reduction and analytical absorption correction \cite{clark1995} were performed with the program suite CrysAlisPro. Using Olex2 \cite{dolomanov2009}, the structure was solved with the SHELXT \cite{sheldrick2015} small molecule structure solution program and refined with the SHELXL2018/3 program package \cite{sheldrick2015crystal} by full-matrix least-squares minimization on F2. PLATON \cite{spek2009} was used to check the result of the X-ray analysis. CCDC 2166026 (for  \ce{Nb4Cu_{0.2}SiSb2}), 2166027 (for  \ce{Nb4Pd_{0.2}SiSb2}), 2166028 (for  \ce{Nb4Pt_{0.14}SiSb2}) and 2166029 (for \ce{Nb4SiSb2}) contain the supplementary crystallographic data for these compounds, and can be obtained free of charge from the Cambridge Crystallographic Data Centre via www.ccdc.cam.ac.uk/data\_request/cif.

Powder X-ray diffraction (PXRD) measurements were performed on a \textit{Rigaku} SmartLab diffractometer using a Cu X-ray source ($K_{\alpha 1} = 1.540600\,$ \AA, $K_{\alpha 2} = 1.544430\,$ \AA) with Cu$K_{\beta}$ filter and collected using a $2\theta $ range of $5\mbox{--}100^{\circ}$. The machine is equipped with a 3 kW sealed X-ray tube, CBO optics and a D/teX Ultra 250 silicon strip detector. Data was recorded using the \textit{SmartLab} Studio II software. \textit{Rietveld} refinements were performed using the \textit{FULLPROF} software package\cite{rodriguez1993recent} and fitting of the diffracted data was done using the \textit{Thompson-Cox-Hastings} pseudo-\textit{Voigt} function with asymmetry correction.\cite{finger1994}

\textit{Physical Properties:} Physical property measurements were carried out on sintered, flat pellets. Temperature-dependent resistivity measurements were performed with a Quantum Design Physical Property Measurement System (PPMS) using a He-3 insert for temperature measurements down to 500 mK. A four-point resistivity measurement method, using silver wires ($50\, \mu m$ diameter) was employed. 

Specific heat measurements were performed from 300 mK to 2 K in a He-3 15 T magnet cryostat with a custom-developed modulated-temperature AC calorimetry technique using an SR830 digital lock-in amplifier, and from 2 - 10 K with a long relaxation technique in a He-4 cryostat. For the latter, each relaxation provides about 1000 data points over a temperature interval of 30-40\% of the base temperature, which has been varied between 2 K and 10 K. The relaxation technique provides a high precision up to 1\% while the AC technique is less accurate but provides high resolutions of $\Delta C / C$ of 10$^{-5}$ at a high density of data points.\cite{mak2013}
Temperature-dependent magnetization measurements were performed using a Quantum Design Magnetic Properties Measurement System (MPMSXL) equipped with a reciprocating sample option (RSO) and a 7 T magnet.

\section{Results and Discussion}

\subsection{Crystal Structure of \ce{Nb4SiSb2}}
\label{Crystal Structure 1}

\begin{table*}
\small
\caption{\ Details of the SXRD measurements and structural refinements for \ce{Nb4SiSb2}, \ce{Nb4Cu_{0.2}SiSb2}, \ce{Nb4Pd_{0.2}SiSb2} and \ce{Nb4Pt_{0.14}SiSb2}}
  \label{tbl_refinement_cryst_data}
  \begin{tabular*}{\textwidth}{@{\extracolsep{\fill}}llllll}
Parameters                                      & \ce{Nb4SiSb2}                & \ce{Nb4Cu_{0.2}SiSb2}          & \ce{Nb4Pd_{0.2}SiSb2}          & \ce{Nb4Pt_{0.14}SiSb2}             \\ 
    \hline                                                                     
Crystal system                                  & tetragonal                   & tetragonal                     & tetragonal                     & tetragonal                         \\
Structure-type                                  & \ce{V4SiSb2}                 & \ce{W5Si3}(defect)             & \ce{W5Si3}(defect)             & \ce{W5Si3}(defect)                         \\
Space group                                     & \textit{I}4/\textit{mcm} (No. 140)    & \textit{I}4/\textit{mcm} (No. 140)      & \textit{I}4/\textit{mcm} (No. 140)      & \textit{I}4/\textit{mcm} (No. 140)          \\
Absorption correction method                    & analytical                   & analytical                     & spherical                      & analytical                         \\
Temperature [K]                                 & 160(1)~                      & 160(1)~ ~                      & 160(1)~ ~                      & 160(1)~ ~                          \\
Lattice parameters [\AA]                        & \textit{a} = 10.3638(2)      & \textit{a} = 10.3954(2)        & \textit{a} = 10.3991(2)        & \textit{a} = 10.3803(2)            \\
                                                & \textit{c} = 4.9151(2)       &\textit{c} = 4.9233(2)          & \textit{c} = 4.93619(16)       & \textit{c} = 4.9348(2)             \\
Cell volume [\AA$^3$]                           & 527.92(3)                    & 532.03(3)                      & 533.81(3)                      & 531.73(3)                          \\
Formula unit/cell                               & 4                            & 4                              & 4                              & 4                                  \\
$\rho_{calcd}$ [g cm$^-3$]                     & 8.093                        & 8.189                          & 8.268                          & 8.376                              \\
$\mu$ [mm$^{-1}$]                               & 149.393                      & 532.03(3)                      & 153.021                        & 155.001                            \\
Crystal size [mm]                               & 0.018 x 0.016 x 0.013        & 0.005 x 0.003 x 0.002          & 0.01 x 0.01 x 0.01             & 0.015 x 0.015 x 0.01               \\
\textit{F}(000)                                 & 1120.0                       & 1143.0                         & 1157.0                         & 1164.0                             \\
Radiation type                                  & Cu $K_{\alpha}$ ($\lambda$ = 1.54184) & Cu $K_{\alpha}$ ($\lambda$ = 1.54184)  & Cu $K_{\alpha}$ ($\lambda$ = 1.54184) & Cu $K_{\alpha}$ ($\lambda$ = 1.54184)        \\                                                                               
2 $\Theta$ range [$^{\circ}$]                   & 12.078 to 146.58             & 12.04 to 148.58                & 12.036 to 147.576              & 12.058 to 147.716                  \\
Index range                                     & \textit{h}[-9,12]            & \textit{h}[-11,9]              & \textit{h}[-11,12]             & \textit{h}[-12,12]                 \\
                                                & \textit{k}[-12,12]           & \textit{k}[-12,12]             & \textit{k}[-12,12]             & \textit{k}[-12,12]                 \\
                                                & \textit{l}[-5,6]             & \textit{l}[-6,6]               & \textit{l}[-6,5]               & \textit{l} [-6,5]     \\
Observed reflections                            & 1466                         & 838                            & 2368                           & 2381                               \\
Independent reflections (2 $\sigma$)          & 165                          & 166                            & 166                            & 167                                \\
$R_{int} $                                      & 0.0278                       & 0.0385                         & 0.0312                         & 0.0298                             \\
$R_{\sigma}$                                    & 0.0127                       & 0.0314                         & 0.0107                         & 0.0117                             \\
Refined parameters                              & 14                           & 16                             & 16                             & 17                                 \\
GOF                                             & 1.363                        & 1.142                          & 1.252                          & 1.240                              \\
$R_1$ (all data) (\%)                           & 1.69                         & 3.33                           & 1.64                           & 1.60                              \\
$wR_1$ ($\geq$ 2$\sigma$) (\%)                  & 1.69                         & 2.96                           & 1.62                           & 1.57                               \\
$wR_2$ (all data) (\%)                          & 4.32                         & 7.53                           & 3.71                           & 3.67                              \\
$wR_2$ ($\geq$ 2$\sigma$) (\%)                  & 4.33                         & 7.37                           & 3.71                           & 3.66                              \\
Max/min residual electron density [e \AA$^{-3}$]& 1.41/-0.94                   & 1.12/-1.52                     & 0.97/-0.98                     & 1.13/-0.85                        \\
\hline
\end{tabular*}
\end{table*}

\begin{table*}
\small
  \caption{\ Atomic coordinates, occupancy, isotropic and anisotropic displacement parameters of the SXRD refinements at 160\, K under atmospheric pressure for the compounds \ce{Nb4SiSb2}, \ce{Nb4Cu_{0.2}SiSb2}, \ce{Nb4Pd_{0.2}SiSb2} and \ce{Nb4Pt_{0.14}SiSb2} (Space Group \textit{I}4/\textit{mcm}, No. 140)}
  \label{tbl_refinement_displacement}
  \begin{tabular*}{\textwidth}{@{\extracolsep{\fill}}lllllllllll}
    \hline
\ce{Nb4SiSb2}           &                          &            &            &            &                         &                       &           &           &             \\
Atom                    & \textit{Wyckoff}         & \textit{x} & \textit{y} & \textit{z} & U(eq) [\AA$^2$]         & U$_{11}$ / U$_{22}$   & U$_{33}$  & U$_{12}$  & Occ.        \\ 
                        & Symbol                   &            &            &            &                         &                       &           &           &             \\
\hline                                                                                                                                                                            
Nb                      & 16\textit{k}             & 0.29305(6) & 0.58530(6) & 1/2        & 0.0111(3)  				& 10.6(4)  / 10.9(4)    & 11.8(4)   & 0.2(2)    & 4.00        \\
Si                      & 4\textit{a}              & 1/2        & 1/2        & 3/4        & 0.0122(3)               & 10.1(11) / 10.1(11)   & 16(2)     & 0         & 1.00        \\
Sb                      & 8\textit{h}              & 0.14037(5) & 0.35963(5) & 1/2        & 0.0119(9)               & 11.8(3)  / 11.8(3)    & 13.0(5)   & -1.5(3)   & 2.00        \\
                        &                          &            &            &            &                         &                       &           &           &             \\
\ce{Nb4Cu_{0.2}SiSb2}   &                          &            &            &            &                         &                       &           &           &             \\
Atom                    & \textit{Wyckoff}         & \textit{x} & \textit{y} & \textit{z} & U(eq) [\AA$^2$]         & U$_{11}$ / U$_{22}$   & U$_{33}$  & U$_{12}$  & Occ.        \\ 
                        & Symbol                   &            &            &            &                         &                       &           &           &             \\
\hline                                                                                                                          
Nb                      & 16\textit{k}             & 0.29297(9) & 0.41603(9) & 1/2        & 0.0080(4)               & 6.2(6)   / 6.4(6)     & 11.4(6)   & 0.3(4)    & 4.00   	  \\
Cu                      & 4\textit{b}              & 0          & 1/2        & 3/2        & 0.021(7)                & 24(8)    / 24(8)      & 16(12)    & 0         & 0.199(16)   \\
Si                      & 4\textit{a}              & 1/2        & 1/2        & 3/2        & 0.0063(12)              & 3.2(17)  / 3.2(17)    & 12(3)     & 0         & 1.00   	  \\
Sb                      & 8\textit{h}              & 0.14385(8) & 0.35615(2) & 1/2        & 0.0110(4)               & 8.8(5)   / 8.8(5)     & 15.3(7)   & 2.3(4)    & 2.00   	  \\
                        &                          &            &            &            &                         &                       &           &           &  			  \\
\ce{Nb4Pd_{0.2}SiSb2}   &                          &            &            &            &                         &                       &           &           &  			  \\
Atom                    & \textit{Wyckoff}         & \textit{x} & \textit{y} & \textit{z} & U(eq) [\AA$^2$]         & U$_{11}$ / U$_{22}$   & U$_{33}$  & U$_{12}$  & Occ.        \\ 
                        & Symbol                   &            &            &            &                         &                       &           &           &  			  \\
\hline                                                                                                                          
Nb                      & 16\textit{k}             & 0.29305(4) & 0.58369(4) & 1/2        & 0.0125(2)               & 12.1(3)  / 12.8(3)    & 12.7(3)   & -0.01(17) & 4.00   	  \\
Pd                      & 4\textit{b}              & 0          & 1/2        & 1/4        & 0.0147(15)              & 13.6(16) / 13.6(16)   & 17(2)     & 0         & 0.199(5)    \\
Si                      & 4\textit{a}              & 1/2        & 1/2        & 3/2        & 0.0131(6)               & 13.1(8)  / 13.1(8)    & 13.0(14)  & 0         & 1.00  	  \\
Sb                      & 8\textit{h}              & 0.14470(4) & 0.35530(4) & 1/2        & 0.0170(2)               & 15.8(2)  / 15.8(2)    & 19.5(3)   & -3.5(2)   & 2.00   	  \\
                        &                          &            &            &            &                         &                       &           &           &   		  \\
\ce{Nb4Pt_{0.14}SiSb2}  &                          &            &            &            &                         &                       &     &                     		  \\
Atom                    & \textit{Wyckoff}         & \textit{x} & \textit{y} & \textit{z} & U(eq) [\AA$^2$]         & U$_{11}$ / U$_{22}$   & U$_{33}$  & U$_{12}$  & Occ.        \\ 
                        & Symbol                   &            &            &            &                         &                       &           &           &   		  \\
\hline                                                                                                                                                                            
Nb                      & 16\textit{k}             & 0.58429(5) & 0.29284(5) & 1/2        & 0.0063(2)               & 6.3(3)   / 6.2(3)     & 6.5(4)    & -0.08(18) & 4.00        \\
Pt                      & 4\textit{b}              & 1/2        & 0          & 3/2        & 0.0122(5)               & 12.5(17) / 12.5(17)   & 12(3)     & 0         & 0.140(3)    \\
Si                      & 4\textit{a}              & 1/2        & 1/2        & 3/2        & 0.0020(6)               & 2.7(8)   / 2.7(8)     & 0.7(16)   & 0         & 1.00        \\
Sb                      & 8\textit{h}              & 0.35697(4) & 0.14303(4) & 1/2        & 0.0101(3)               & 9.0(3)   / 9.0(3)     & 12.5(4)   & -3.1(2)   & 2.00        \\
                        &                          &            &            &            &                         &                       &           &           &    		  \\
\hline
\end{tabular*}
\end{table*}

In Figure \ref{fig_nb4sisb2_sc_and_rietveld}a\ and b, we present the crystal structure and the unit cell of the single-crystal refinement of \ce{Nb4SiSb2}, shown along the \textit{c}-direction and along the \textit{b}-direction, respectively. The structure of \ce{Nb4SiSb2} was determined by means of single crystal X-Ray diffraction (SXRD) at 160\, K and the elemental composition was confirmed using EDX analysis at ambient temperature (Supplementary Information). 

We find \ce{Nb4SiSb2} to crystallize in the tetragonal space group \textit{I}4\textit{/mcm} with the lattice parameters \textit{a}~= \textit{b} = 10.3638(2)\, \AA~ and \textit{c}~= 4.9151(2)\, \AA~ with the corresponding calculated cell volume of \textit{V} = 527.92(3)\, \AA$^3$. Hence, it is found to adopt the same centrosymmetric structure type that was previously reported for \ce{V4SiSb2}\cite{Wollesen1996, Richter1997}. The crystallographic data and the details of the structure refinement are summarised in Table\, \ref{tbl_refinement_cryst_data}. All crystallographic positions as well as the anisotropic displacement parameters are presented in Table \ref{tbl_refinement_displacement}. 

In the structure of \ce{Nb4SiSb2} each atom occupies one atomic site: The niobium atoms are located at the 16\textit{k} \textit{Wyckoff} position, silicon occupies the 4\textit{a} and antimony the 8\textit{h} \textit{Wyckoff} positions. Silicon forms thereby columns which can be interpreted as $_{\infty}$[Nb$_{8/2}$Si] chains along the \textit{c}-direction as shown in Figure \ref{fig_nb4sisb2_sc_and_rietveld}(c). The Si--Si bonding distance in \ce{Nb4SiSb2} within the columns is 2.4576(1)\, \AA, which is in good agreement with the ones found in \ce{V4SiSb2}\cite{Wollesen1996} and comparable to Si--Si bond distances in similar structures.\cite{Badding1990, Wang2001} Each Si atom is surrounded by eight Nb atoms with a distance of 2.6252(6)~\AA\, forming antiprisms with the surrounding neighbour atoms. Nb has a coordination number (CN) of 13 consisting of six Nb neighbours located in the $_{\infty}$[Nb$_{8/2}$Si] column, one Nb in the adjacent $_{\infty}$[Nb$_{8/2}$Si] column, two Si, and four Sb neighbours located in between the two columns. The Nb--Nb distances range from 3.0275(8) to 3.2807(9)~\AA. These distances, together with the relatively short intercolumn distance between two Nb atoms of 3.0449(13)~\AA~are in good agreement with distances found in comparable structures.\cite{Janger1975,Lomnytska2006}  
Also, the Nb--Sb distance ranging from 2.8238(7)~\AA~to 2.9781(4)~\AA~is in good agreement with the distances found in the related compounds, such as e.g. in \ce{Nb5Sb4}.\cite{Lomnytska2006} Each Sb has eight Nb neighbours and therefore a CN of 8. Another feature of this structure are the voids at the 4\textit{b} \textit{Wyckoff} position. These void positions are surrounded by four Sb atoms. These form void channels along the \textit{c}-direction. If these void positions were fully occupied, then the \ce{V4SiSb2} structure would be equivalent to the \ce{W5Si3} structure.\cite{Wollesen1996}

The validity of the structural model, the phase purity, and the homogeneity of the sample were confirmed by means of PXRD at ambient temperature and SXRD at 160\,K. The reliability factors of the SXRD refinement can be found in the supplementary information. In figure \ref{fig_crystal_structures_nb4msisb2}(d) the PXRD pattern of the polycrystalline sample is shown, with its respective \textit{Rietveld} refinement. We find the lattice parameters of \textit{a}~= \textit{b}~= 10.3686(4)~\AA~, and \textit{c}~= 4.9193(2)~\AA~, as well as a calculated cell volume of \textit{V} = 528.86(3) \AA$^3$. Hence, the SXRD and PXRD refinements and structural solutions are in excellent agreement with each other (Supplementary Information).

\begin{figure}[h]
\centering
  \includegraphics[width =0.9\linewidth]{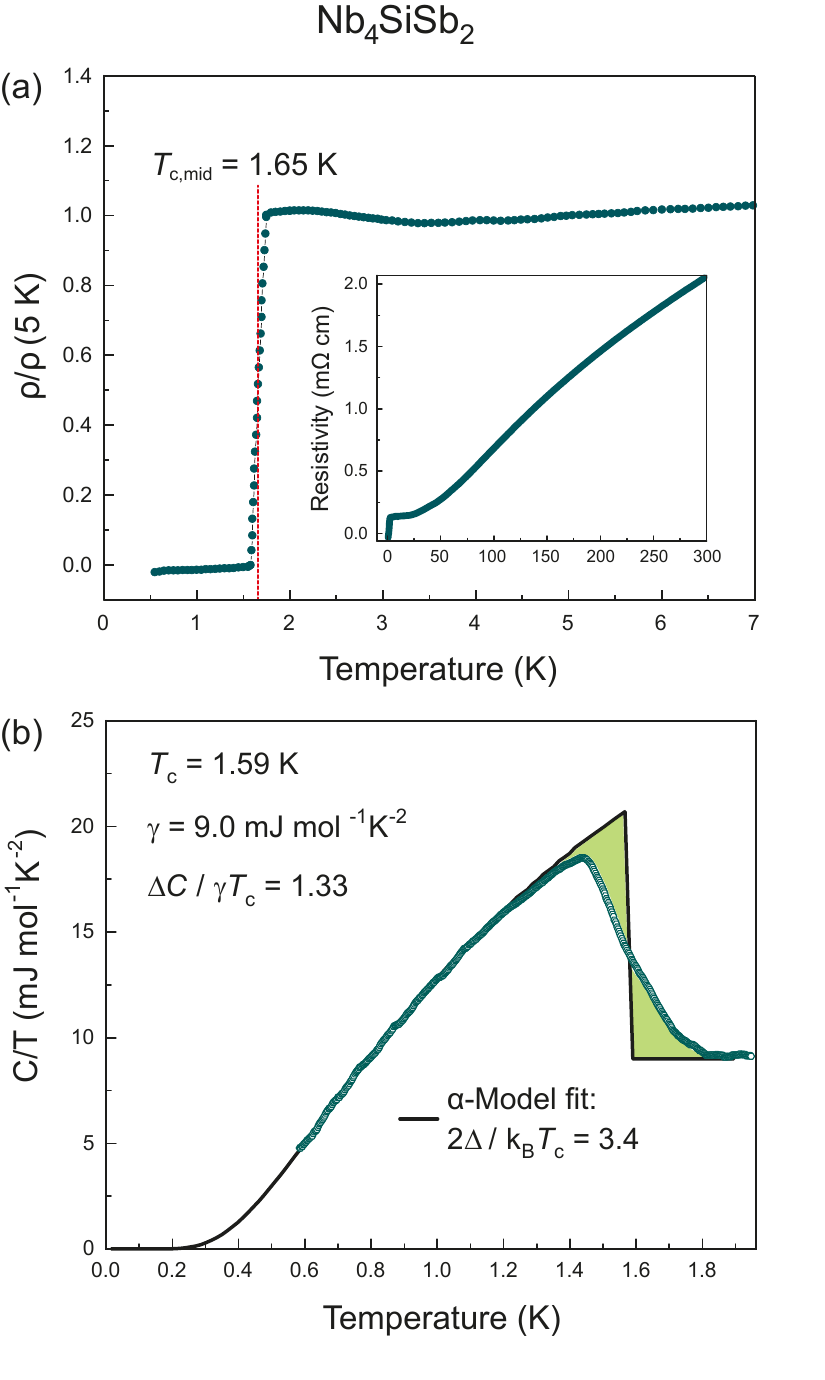}
  \caption{(a) normalized, low-temperature resistivity of \ce{Nb4SiSb2} in a temperature range between $T$ = 500\, mK and 5\, K measured in zero field $\mu_0H = 0\, T$. Inset: Temperature-dependant resistivity of \ce{Nb4SiSb2} in zero field $\mu_0H = 0\, T$ as $\rho(T)$ between T = 500 mK and 300 K (b) Specific heat capacity of \ce{Nb4SiSb2} in a temperature range between $T$ = 580 mK and 2\, K. The black line corresponds to a fit using the $\alpha$-model.}
  \label{fig_temR_tempSH}
\end{figure}

\subsection{Superconducting Properties of \ce{Nb4SiSb2}}
\label{Superconducting Properties 1}

In Figure \ref{fig_temR_tempSH}(a), we show the temperature-dependant resistivity of \ce{Nb4SiSb2} in zero field $\mu_0H = 0\, T$ as $\rho(T)$ between T = 300 K and 500 mK (inset) and in the vicinity of the superconducting transition. A sharp drop in the resistivity is observed at low temperature, corresponding to a transition to a superconducting state. The transition midpoint of $T_{\rm c,mid}\approx1.65$ K and reaches a state of zero resistance at $T_{\rm zero}\approx 1.56$ K. The transition is comparably sharp with a transition width of $\Delta T = 0.18$ K in the resistivity. The residual resistivity $\rho$(1.8 K) = 0.14~m\,$ \Omega\,$ cm at 1.8 K and the room temperature resistivity value of $\rho(300\,$ K) = 2.06~m\,$ \Omega\,$ cm, result in a residual resistivity ratio (RRR) here defined as RRR = $\rho(300\,  K)/\rho(1.8\, K) = 14.96$. This RRR value corresponds to the value of a good metal.

The bulk nature of the superconductivity in \ce{Nb4SiSb2} is confirmed by low-temperature specific-heat measurements. Temperature-dependent specific-heat measurements are of particular importance to prove the bulk nature of a superconductor.\cite{Carnicom2019,Witteveen2021} 

In Figure \ref{fig_temR_tempSH}(b), we present the temperature-dependent specific heat $C(T)/T$ of \ce{Nb4SiSb2} in a temperature range between $T =$ 600\, mK and 2\, K. We find a clearly pronounced discontinuity in the specific heat, resulting from the superconducting transition. The data was fitted using the $\alpha$-model.\cite{Padamsee1973,Johnston2013} Thereby, an entropy conserving construction was used to determine the critical temperature, $T_{\rm c} \approx$  1.6\, K. This value is in good agreement with the critical temperature from the resistivity measurement. From the $\alpha$-model fit, we obtained $\alpha = 1.7$ and the Sommerfeld constant of $\gamma$ = 9.00~mJ\, mol$^{-1}$\, K$^{-2}$. We find a ratio for the normalized specific-heat jump of $\Delta$ C/ $\gamma$ $T_{\rm c}$ = 1.33\, mJ\, mol$^{-1}$\, K$^{-2}$, which confirms the bulk nature of the superconductivity, as this value is close to the weak-coupling BCS ratio of 1.43. This corresponds to a value of the superconducting gap of $2 \Delta(0) = 3.4\, k_{\rm B} \ T_{\rm c}$.

\begin{figure}[h]
 \centering
 \includegraphics[width =0.95\linewidth]{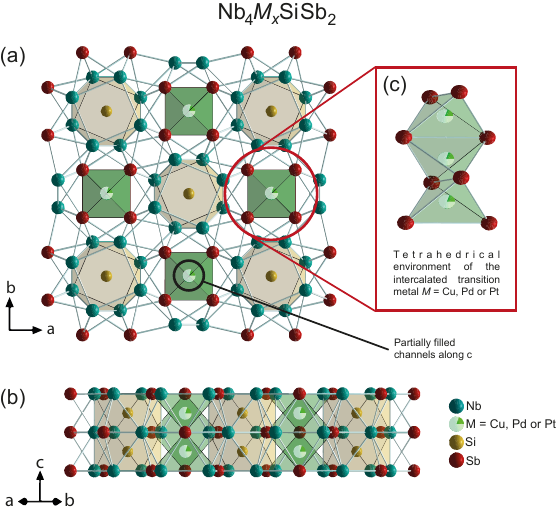}
 \caption{Schematic representation of the structures of \ce{Nb4\textit{M}_xSiSb2} from (a) the \textit{c}-direction and (b) a high symmetry-direction. (c) illustrates the intercalated transition metal \textit{M} (Cu, Pd, or Pt) in its environment within the $_{\infty}$[Sb$_{4/2}$\textit{M}$_{x}$] channels. The crystal structures were obtained from SXRD refinements.}
 \label{fig_crystal_structures_nb4msisb2}
\end{figure}

Under the assumption of a degenerate electron gas of non-interacting particles, the electronic contribution to the heat capacity in a solid at low temperatures is proportional to the density of states at the \textit{Fermi} level $D(E_{\rm F})$ and linear in $T$. With the previously determined value of $\gamma$ = 9.00~mJ\, mol$^{-1}$\, K$^{-2}$, the density of states at the \textit{Fermi} level can be calculated as described by \textit{F. Heiniger et al.}\cite{Heiniger1966} according to

\begin{equation}
\label{eq_DEF}
  C_{el} = \gamma\, T = \frac{\pi^2}{3}\, k_B^2\, D(E_F)T.
\end{equation}

We obtain for \ce{Nb4SiSb2} a density of states at the \textit{Fermi} level of $D(E_F) = 3.8$ states~eV$^{-1}$. 

Magnetic susceptibility measurements of \ce{Nb4SiSb2} were conducted in the normal-state, i.e. in a temperature range between $T$ = 10 K to 300 K, in an external field of $\mu_0H = 1$ T. The observed temperature-independent positive magnetic moment corresponds to a \textit{Pauli}-paramagnet (see Supplementary Information). A summary of all obtained physical parameters can be found in Table \ref{tbl_summary_phys_data}.

\begin{table*}
\small
  \caption{\ Summary of the physical parameters for \ce{Nb4SiSb2}, \ce{Nb4Cu_{0.2}SiSb2}, \ce{Nb4Pd_{0.2}SiSb2} and \ce{Nb4Pt_{0.14}SiSb2}}
  \label{tbl_summary_phys_data}
  \begin{tabular*}{\textwidth}{@{\extracolsep{\fill}}llllll}
    \hline
Parameter                           & Units                         & \ce{Nb4SiSb2}                         & \ce{Nb4Cu_{0.2}SiSb2}          & \ce{Nb4Pd_{0.2}SiSb2}         & \ce{Nb4Pt_{0.14}SiSb2}           \\
\hline                                                                                      
$T_{\rm c,resistivity}$             & K                             & 1.65                                  & 1.16                           & 0.76                          & 0.84                             \\ 
$T_{\rm c,specificheat}$            & K                             & 1.59                                  & -                              & -                             & -                                \\
RRR                                 & -                             & 14.96                                 & 4.54                           & 1.56                           & 1.70                             \\
$\rho(300)$                         & mJ\, $\Omega$\, cm              & 2.06                                  & 0.70                           & 8.46                          & 2.49                             \\
$\rho_0$                            & mJ\, $\Omega$\, cm              & 0.13                                  & 0.15                           & 5.43                          & 1.46                             \\
Type of magnetism                   & -                             & \textit{Pauli}-paramagnetic           & \textit{Pauli}-paramagnetic    & \textit{Pauli}-paramagnetic   & \textit{Pauli}-paramagnetic       \\   
$\gamma$                            & mJ\, mol$^{-1}$\, K$^{-2}$    & 9.00                                  & 7.5                            & - &                                                               \\
$\Delta\, C / T_{\rm c} \gamma$     & -                             & 1.33                                  & 1.2                            & - &                                                               \\
$2 \Delta(0)$                       & meV                           & 0.47                                  & 12                             & - &                                                               \\
$D(E_F)$                            & states eV$^{-1}$ per f.u.     & 3.82                                  & 3.18                           & - &                                                                \\
\hline
\end{tabular*}
\end{table*}

\begin{figure}
 \centering
 \includegraphics[height=13cm]{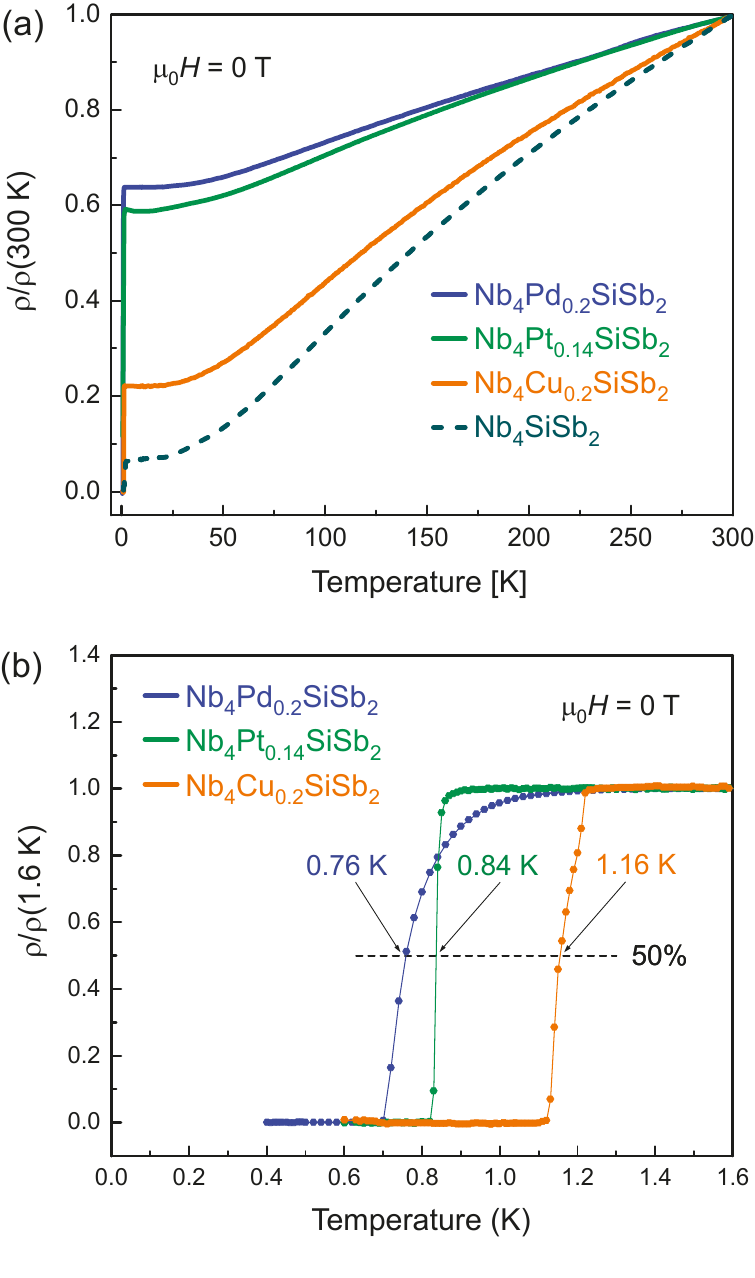}
 \caption{(a) Temperature-dependent resistivity of \ce{Nb4Cu_{0.2}SiSb2}, \ce{Nb4Pd_{0.2}SiSb2} and \ce{Nb4Pt_{0.14}SiSb2} (a) over the whole temperature range between $T$ = 400\, mK and 300\, K, and (b) in the vicinity of the superconducting transitions at low temperatures. All measurements were performed in  zero field $\mu_0H = 0\, T$.}
 \label{Plot_R_over_R2_CuPdPt}
\end{figure}

\subsection{Crystal Structures of \ce{Nb4Cu_{0.2}SiSb2}, \ce{Nb4Pd_{0.2}SiSb2}, and \ce{Nb4Pt_{0.14}SiSb2}}
\label{Crystal Structure 2}

We have synthesized the three compounds \ce{Nb4Cu_{0.2}SiSb2}, \ce{Nb4Pd_{0.2}SiSb2}, and \ce{Nb4Pt_{0.14}SiSb2}. Here, the void 4\textit{b} Wyckoff positions in \ce{Nb4SiSb2} are partially filled with a transition metal \textit{M} = Cu, Pd, or Pt, respectively. In Figure \ref{fig_crystal_structures_nb4msisb2}, we show a schematic representation of the unit cell along the \textit{c}-direction and a high symmetry-direction of \ce{Nb4\textit{M}_{x}SiSb2}, where \textit{M} = Cu, Pt and Pd with $x$ = 0.2, 0.14 and 0.2. The crystal structures of all three compounds were determined using SXRD at 160\, K and PXRD  diffraction at room temperature.

All samples were found to be single phase by means of PXRD measurements and corresponding Rietveld refinements (Supplementary Information). Atomic compositions were confirmed using EDX analysis (Supplementary Information). 

All three structures are in good agreement with the previously reported structure for \ce{Nb4Pd_{0.5}\textit{Z}Sb2} with \textit{Z} = Cr, Fe, Co, Ni, Si, where it was thought that a half-occupied Pd 4\textit{b} site was necessary to stabilize these compounds.\cite{Wang2001} In contrary to this previous assumption, we found here that the channels were in our case either unoccupied or filled with 0.2 or 0.14 respectively (in case of Pt), independent of the initially used starting stoichiometry. These results indicate that, with improved synthesis methodologies, the continuous solid solution might be accessible in the future. All information regarding the lattice parameters, crystallographic data, and details of the structure refinements are summarized in Table \ref{tbl_refinement_cryst_data}. 

\subsection{Electronic Properties of \ce{Nb4Cu_{0.2}SiSb2}, \ce{Nb4Pd_{0.2}SiSb2} and \ce{Nb4Pt_{0.14}SiSb2}}
\label{Superconducting Properties 2}

In Figure\, \ref{Plot_R_over_R2_CuPdPt} we present the temperature-dependent resistivity and the normalized low-temperature resistivity $\rho(T) / \rho(1.6\, K)$ in a temperature range between $T$ = 400\, mK and 1.6\, K for \ce{Nb4Cu_{0.2}SiSb2}, \ce{Nb4Pd_{0.2}SiSb2} and \ce{Nb4Pt_{0.14}SiSb2}, measured in zero field $\mu_0H = 0\, T$. 

We find all three compounds to undergo a transition to a superconducting state at low temperatures. The critical temperature midpoints are determined as $T_{\rm c,mid} \approx 1.16$ K for \ce{Nb4Cu_{0.2}SiSb2}, $T_{\rm c,mid} \approx 0.76$ K for \ce{Nb4Pd_{0.2}SiSb2} and $T_{\rm c,mid} \approx 0.84$ K for \ce{Nb4Pt_{0.14}SiSb2}. All three compounds with atoms in the void position of \ce{Nb4SiSb2} have lower transition temperatures than the parent compound.

For comparison, we have performed specific heat measurements in the normal state of \ce{Nb4SiSb2} and \ce{Nb4Pt_{0.14}SiSb2} (shown in the SI). For \ce{Nb4SiSb2} we find values for $\gamma_n$ and $\beta$ of 8.40\, mJ\, mol$^{-1}$\, K$^{-2}$  and 0.16\, mJ\, mol$^{-1}$\, K$^{-4}$, respectively. The $\gamma_n$ value of this fit is in good agreement with the more accurate low-temperature value discussed above. For \ce{Nb4Pt_{0.14}SiSb2} we find values for $\gamma_n$ and $\beta$ of 9.10\, mJ\, mol$^{-1}$\, K$^{-2}$ and 0.31\, mJ\, mol$^{-1}$\, K$^{-4}$, respectively. We note that the values for $\gamma_n$  differ only slightly, indicating a small change of the electronic properties upon void position filling. We find, however, that $\beta$ changes quite strongly. These findings indicate that the decrease of the superconducting transition temperature is likely caused by a change in the phonons, and the vibrations, respectively. 

\ce{Nb4Pd_{0.2}SiSb2} has the lowest critical temperature of the doped compounds, as well as the lowest RRR value of RRR = $\rho(300\, K) / \rho(1.8\, K) = 1.56 $. 
\ce{Nb4Cu_{0.2}SiSb2} with RRR = 4.54 and \ce{Nb4Pt_{0.14}SiSb2} with RRR = 1.70 follow the descending trend observed for the critical temperatures accordingly. These low RRR values correspond to a poor metallic behaviour and are 3 to 24 times smaller than the RRR of the parent compound \ce{Nb4SiSb2}. The pronounced effect on the physical properties on void position doping becomes clearly apparent in the large change of the RRR values. The nature of the change is, however, not only affected by the electronic states, but also by the phonons and by impurity state scattering. 

\section{Conclusion}
\label{Conclusion}

We have reported on the discovery of the centrosymmetric structure compound \ce{Nb4SiSb2}. This phase was found to crystallize in the tetragonal \ce{V4SiSb2}-type structure. We found \ce{Nb4SiSb2} to undergo a transition to a superconducting state at a critical temperature of $T_{\rm c} \approx\, 1.6\, K$. The bulk nature of the superconducting transition was evidenced by a clear discontinuity in specific heat, with a normalized specific heat jump of $\Delta C(T_{\rm c})/\gamma T_{\rm c} =\, $1.33\, mJ\, mol$^{-1}$\, K$^{-2}$, close to the weak-coupling BCS value. Furthermore, we have shown that the  unoccupied 4\textit{b} Wyckoff site in \ce{Nb4SiSb2} can be partially occupied with the transition metals Cu, Pd, or Pt. 

These compounds crystallize in a tetragonal variant of the \ce{W5Si3}-type structure with partially occupied channels, extending along the \textit{c}-direction. All three compounds were found to be superconductors with transitions temperatures of $T_{\rm c} \approx\, 1.2\, K$ for \ce{Nb4Cu_{0.2}SiSb2}, $T_{\rm c} \approx\, 0.8\, K$ for \ce{Nb4Pd_{0.2}SiSb2} and $T_{\rm c} \approx\, 0.8\, K$ for \ce{Nb4Pt_{0.14}SiSb2}. We find that the insertion of a host atom into the void positions strongly affects the electronic and superconducting properties of this material. 

Hence, our results indicate that this and related compounds might be promising host structures for the discovery of new superconducting materials, as they allow for a controlled manipulation of the electronic and phononic properties by chemical manipulation. 


\section*{Conflicts of Interest}
There are no conflicts to declare.


\section*{Acknowledgements}
This work was supported by the Swiss National Science Foundation under Grant No. PCEFP2\_194183 and by grants from the Research Grants Council of the Hong Kong Special Administrative Region, China (GRF-16302018 \& C6025-19G-A).

\balance

\bibliography{rsc} 
\bibliographystyle{rsc} 

\end{document}